# Big Data as a Mediator in Science Teaching: A Proposal

Renato P. dos Santos[1]

[1] ULBRA/PPGECIM, renatopsantos@ulbra.edu.br

**Abstract**

We live in a digital world that, in 2010, crossed the mark of one zettabyte data. This huge amount of data processed on computers extremely fast with optimized techniques allows one to find insights in new and emerging types of data and content and to answer questions that were previously considered beyond reach. This is the idea of Big Data. Google now offers the Google Correlate analysis public tool that, from a search term or a series of temporal or regional data, provides a list of queries on Google whose frequencies follow patterns that best correlate with the data, according to the Pearson determination coefficient $R^2$. Of course, "correlation does not imply causation." We believe, however, that there is potential for these big data tools to find unexpected correlations that may serve as clues to interesting phenomena, from the pedagogical and even scientific point of view. As far as we know, this is the first proposal for the use of Big Data in Science Teaching, of constructionist character, taking as mediators the computer and the public and free tools such as Google Correlate. It also has an epistemological bias, not being merely training in computational infrastructure or predictive analytics, but aiming at providing students a better understanding of physical concepts, such as phenomena, observation, measurement, physical laws, theory, and causality. With it, they would be able to become good Big Data specialists, the so needed 'data scientists' to solve the challenges of Big Data.

## Introduction

The digital universe we live in crossed the mark of one zettabyte (approximately $10^{21}$) data (Zikopoulos et al., 2013, p. 9), made up mainly of postings and 'likes' in social networks, phone calls, SMS messages, images and videos on mobile phones uploaded to YouTube, digital high-definition movies, banking data swiped in an ATM, security footage at airports and major events, subatomic collisions recorded by the Large Hadron Collider at CERN (Gantz and Reinsel, 2012). We live in the Big Data world.

Among the many available definitions of Big Data, we prefer the following, for we judge it the most insightful for the purposes of this study:

> Big data is more than simply a matter of size; it is an opportunity to find insights in new and **emerging** [emphasis added] types of data and content, to make your business more agile, and to answer questions that were previously considered beyond your reach. Until now, there was no practical way to harvest this opportunity. Today, IBM's platform for big data uses state of the art technologies including patented advanced analytics to open the door to a world of possibilities. (IBM, n.d.).

We believe, as Searls (2013), that, as it happened with PCs, Internet and mobile communication, we as individuals will be able to do more with our own data than big enterprises can, without the need of learning Hadoop[1], MapReduce[2] or many others, as soon as the means to do so are available. In fact, we see the coming out of very powerful yet affordable Big Data tools, such as *BigSheets*[3], *Tableau*[4], *Karmasphere*[5], *Revolution Analytics*[6], and *HDInsight*[7].

According to Mattmann (2013), a new breed of 'data scientists' to solve Big Data challenges is necessary. For Kate Mueller (Dumbill et al., 2013), to say that only computer science people make good Big Data experts or good data analysts is a mistake. Mattmann (2013) insists that natural scientists, too, should become familiar with Big Data.

---

[1] Hadoop is a software platform written in Java for storage and large-scale processing of data-sets on clusters. In *Wikipedia, The Free Encyclopedia*.
[2] MapReduce is a programming model for processing large data sets with a parallel, distributed algorithm on a cluster. In *Wikipedia, The Free Encyclopedia*.
[3] http://www-01.ibm.com/software/ebusiness/jstart/bigsheets/
[4] http://www.tableausoftware.com/
[5] http://karmasphere.com/what-we-do
[6] http://www.revolutionanalytics.com/
[7] http://www.windowsazure.com/pt-br/home/features/hdinsight/

The main objective of this project is to investigate the feasibility of using Big Data in Science Teaching, taking as mediators, the computer and public and free Big Data tools such as *Microsoft Power Map*[8] (previously known as *GeoFlow*), *Google Trends*[9], *Google Correlate*[10], and other that shall arise soon. It aims at providing students a better understanding of physical concepts, such as phenomena, observation, measurement, physical laws, theory, and causality, and, therefore, being able to become good Big Data specialists, the so needed 'data scientists' to solve the challenges of Big Data.

**Our proposal**

In our view, the central parameter that distinguishes Big Date from previous processes of data analysis is not the so emphasized huge size of databases, but the emergence of new facts and unsuspected correlations. Here we use the concept of (weakly) emergence in the sense of a new, unexpected, non-obvious property of a complex system that is distinct from the properties of the different parts of the system, while being deductible from and caused by them (cf. Chalmers, 2006).

We do not expect strong emergent phenomena to arise from Big Data as their existence would imply in new fundamental laws of Nature needed to explain them (Chalmers, 2006). As a matter of fact, we are counting on the deductibility aspect of the weakly emergent correlations for students to be able to seek scientific explanations for them.

We do not foresee reasons for a decrease in the production of digital data. Thus, even if Big Data fashion is replaced by the next big thing, we are sure Science and Economics will require some sort of analysis of these huge masses of data. From our experience, we observed that producing questions is more challenging than getting answers to them with these Big Data tools. Here, we believe, is where Science professionals are most needed.

---

8  http://www.microsoft.com/en-us/download/details.aspx?id=38395
9  http://www.google.com/trends/
10 http://www.google.com/trends/correlate

Our proposal is based on Papert's constructionism (1985), which emphasizes the importance of tools, media, and contexts in human development and how his dialogues with artifacts promote self-learning and facilitate the construction of new knowledge (Ackermann, 2001).

It is important, however, to understand that Papert's microworlds are not mere learning objects, from which students learn, but intellectual environment in which the emphasis is on the process (Papert, 1980, p. 184). That was his controversial vision of computers as learning tools rather than teaching tools (1980, p. 19), anticipating Rosa's idea (2008) of students *to-think-with* and *to-learn-with* the computer. It is also worth mentioning Turkle's view (2007) of computers as evocative objects, that is, *things we think with*.

Among previous proposals of using Big Data as a tool for teaching and learning, we have to mention (Baram-Tsabari and Segev, 2009a, 2009b, in press; Segev and Baram-Tsabari, 2012) that propose to use *Google Trends*, *Google Zeitgeist*, and *Google Insights for Search* for research and discussion on the public understanding of science and the distinction between science and pseudoscience in the classroom. Bülbül (2009) and Yin et al. (2013) propose to determine and discuss trends in Physics and Education through keyword searches via *Google*, *Google Scholar*, and *Google Trends*.

Our proposal is distinct from those ones and is based on *Google Correlate*, a newer tool which will be discussed below.

As it is known, the Google search engine not only performs alleged one hundred billion monthly searches of Web terms (Sullivan, 2012), but it stores them all, identified by time and place of origin on its huge data centers around the world. This information is subsequently used by the advertising programs managed by Google, such as *DoubleClick*, *Google Analytics*, *Google AdWords,* and *Google AdSense*, from which comes more than 90% of the Google company income (Google Inc., 2013). Google Inc. has as well released various public analytical tools making use of this stored information.

*Google Trends* made a discrete apparition as a mere feature in Google search engine on December 20, 2005 and was officially launched only on May 10, 2006 (Mayer, 2006), built on the idea behind the *Google Zeitgeist* (Mayer, 2006) (closed 22 May 2007 and replaced by *Hot Trends,* a dynamic feature in *Google Trends*). It allows users to sort through several years of Google search queries from around the world to get a graphical plotting showing the popularity of particular search terms over time.

Since May 2011, the Google company offers the public the *Google Correlate* analysis tool, built on the previous work on *Google Trends*, *Google Insights for Search* (merged into *Google Trends* on Sep. 2012), and *Google Flu Trends*[11] (Mohebbi et al., 2011). According to those authors, *Google Correlate* employs a novel approximate nearest neighbor (ANN) algorithm that allows for automated query selection across millions of candidate queries in an online search tree for any temporal or spatial pattern of interest. According to the authors, it produces results similar to the batch-based approach employed by *Flu Trends* but in a fraction of a second (Mohebbi et al., 2011). Initially, only US data was available, but on January 3rd, 2012, it was expanded to 49 additional countries (Mohebbi, 2012).

*Google Correlate* takes an input, which can be:

- an individual search term,
- a temporal data series uploaded by the user,
- a spatial data series uploaded by the user, or
- a sketch of a graph made with its 'Search by Drawing' feature.

It then finds, from millions of possible candidates through an automated process, the set of individual search queries whose spatial or temporal pattern are most highly correlated (measured by Pearson correlation coefficient $R^2$) with the input (Mohebbi et al., 2011).

---

11 http://www.google.org/flutrends/

According to Mohebbi et al. (2011), Google Correlate requires inputs with unique spatial or temporal patterns. Inputs with little variation or very regular variation are unlikely to surface meaningful results. However, inputs with a unique variation may still not lead to any result due to a lack of information-seeking behavior for it.

A number of academic works based on *Google Correlate* can be found in the scientific literature in several domains, including Public health, Economics, Sociology, and Meteorology, among others.

Of course, one must be aware that the underlying cause of search behavior can never be known, that this correspondence may not hold in the future due to changes in user behavior, which are unrelated to the target behavior, and that Google users do not represent a random sample of the population (Mohebbi et al., 2011). In fact, on 2013, Google Flu Trends predicted a flu outbreak with nearly double the intensity reported by U.S. authorities, possibly caused by a feedback process in Google forecasts (Butler, 2013).

On the other hand, as pointed by Lazer et al. (2014), Google and its derivatives *Google Trends* and *Google Correlate* are constantly changing their algorithms for numerous technical, strategic, and commercial reasons. Moreover, this change prevents the replication of results and consequently compromises the empirical research of the scientists who use these data sources over time and casting doubt if they are watching robust patterns or trends evanescent. As discussed below, however, our proposal will be less affected by these issues because it is based on the qualitative aspects of the correlation and not on the quantitative ones.

It is also worth remembering the statisticians' warning "correlation does not imply causation" (Field, 2003, p. 10), meaning that establishing a correlation does not imply a causal relationship, because we do not know what caused what. However, "correlation is not causation, but it sure is a clue" (Tufte, 2006, p. 5). It is according to this view that we believe that there is a potential for these Big Data tools to find unexpected, and even unusual, correlations which may, however, serve as clues to interesting phenomena from the pedagogical and even scientific point of view.

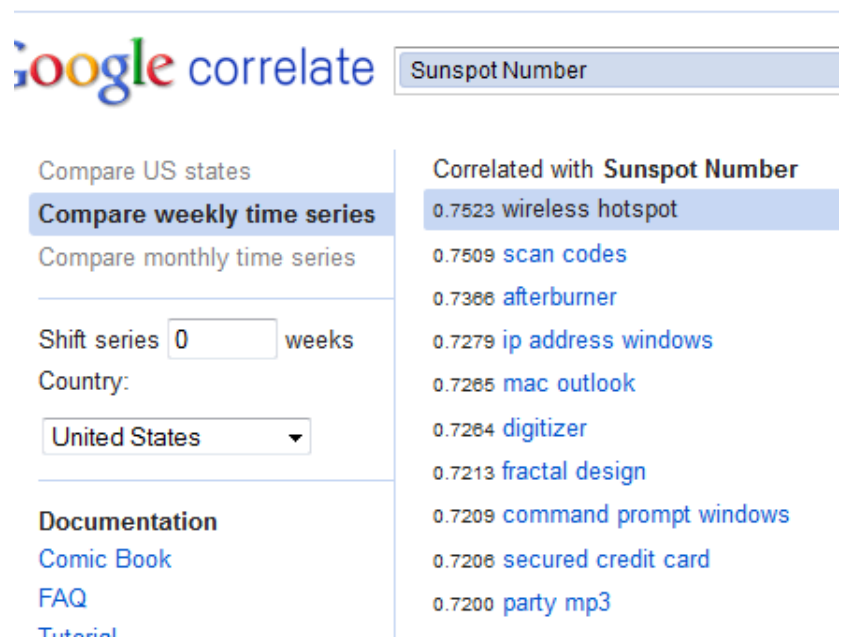

**Figure 1** - Search term frequencies on Google positively correlated with the weekly change in the sunspot number from January 5, 2003 to March 31, 2013.
Source: *Google Correlate* (http://www.google.com/trends/correlate).

As an example, we used the weekly variation of solar activity, measured by the change in the sunspot number[12] from January 5, 2003 to March 31, 2013. One observes, from Figure 1, a good correlation for several terms, and the best correlated ($R^2 = 0,7523$) was 'wireless hotspot,' which means locations were wi-fi is freely available.

In the graph produced by *Google Correlate* for the term 'wireless hotspot' (Figure 2), this correlation is fairly apparent.

---

[12] Data obtained from http://www.ngdc.noaa.gov/stp/space-weather/solar-data/solar-indices/sunspot-numbers/international/listings/

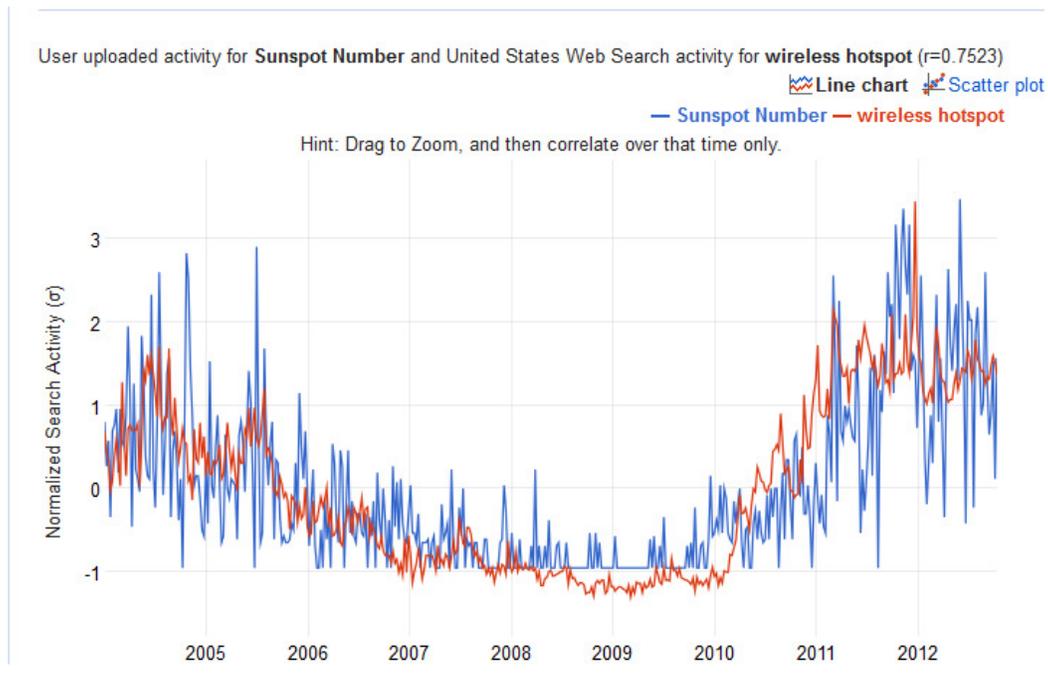

**Figure 2** - Comparison between the frequency of the search term 'wireless hotspot' in Google and the weekly change in the sunspot number from January 5, 2003 to March 31, 2013.

Initially, one may not see a causal relationship between searches for 'wireless hotspots' on Google and variations in solar activity. However, it is known that solar activity has several effects on our daily life, among which variations in radio-communication conditions, geomagnetic storms and disturbances, changes in climatic conditions, and polar auroras (Vitinskii, 1965). Thus, a possible causal mechanism for the observed correlation would be that this activity maximum hinders the reach of hotspots and, therefore, users that were accustomed to using certain hotspots would be forced to look for new hotspots to connect.

By our proposal, this is a fruitful learning moment for the science student: observed a new, unexpected emergent phenomenon (correlation), he should seek a scientific explanation (causation) for it. To confirm or refute the above outlined hypothesis, the students would need to deepen their research from various other sources, an effort which would be extremely productive in terms of science learning. Here we see the Big Data tools acting as mediators in science teaching, with the computers as evocative objects (Turkle, 2007), the students thinking and learning *with-the-computer* (Rosa, 2008).

In our understanding, this is an example of a Big Data weakly emergence (cf. Chalmers, 2006): a new, unexpected, non-obvious property (the correlation between 'sunspot number' and 'wireless hotspot' search term frequency variations) of a complex system (Google users' seeking behavior) that is distinct from the properties of the different parts of the system (individual Google users, unique Google searches), while being deductible from and caused by them, as by the argument above raised.

Is worth emphasizing that our proposal is based on the search for the scientific explanation for the correlation, not on the discussion about its quantitative aspects. The emphasis is on the search process of possible causes for the correlation and not the identification of the correlation itself. With that, it should be less affected by temporal aspects of the correlations' non-replicability, as pointed out earlier.

From the earlier discussion, one sees that this proposal has an epistemological bias. More than merely training students in computer infrastructures or predictive analysis, the goal of the teaching activities to be performed during this project is to provide students, future Science professionals, a better understanding of the construction of physical knowledge especially of the phenomenon, observation, measurement, physical laws, theory, and causality concepts, among others. Furthermore, it is expected they gain a preparation, both in technical and ethical terms, for the scientific challenges posed by Big Data to the real world in which they will exercise their professions.

**Materials and Methods**

This proposal is being developed within the 'History and Epistemology of Physics' course, from the pre-service Physics teacher training program at Ulbra - Lutheran University of Brazil, which has this researcher as the lecturer, and is being held during the first school semester 2014.

The objectives of this course are:

> To present the student to the evolution of physical thinking. [...] To provide the student with elements of Epistemology to help him understand what is the reality for Physics, **what is Law, Theory** [emphasis added], Science, Pseudoscience, **Empiricism, Inductivism** [emphasis added], Rationalism, Determinism, Positivism, Critical Rationalism, [... ], among other key topics (Ulbra, 2014, p. 1).

Initially, at the end of the class in which the concepts of phenomenon, observation, measurement, physical laws, theory, and causality were discussed, it was presented the concept of Big Data as well as a brief epistemological vision about its potential in Physics teaching and scientific research, its main public tools and their use, and the proposal of their application activities in the course.

Students were then asked to search freely for search terms correlations related to Physics Teaching in Google Correlate, as done in the above example. Once the best correlated search terms were obtained, the students are now deepening their research on these correlations from several other sources, looking for possible scientific explanations (causation) for them.

As a final activity for the course partial evaluation, they shall present their research results in the form of seminars to the rest of the class. Furthermore, they were informed that the best research works will be converted into scientific articles on science teaching.

At the end of the semester, participants will be interviewed about their perception of the activity in general and it will be made an evaluation of the development of their understanding of the worked physical concepts through concept maps (Novak and Gowin, 1984).

These activities are being assisted by an undergraduate research student of this same course, who is in charge of helping students in their difficulties and recording their comments and questions for further investigation. A careful record of the activities and reactions of the students is also being done.

At the end of the course, on July 2014, a critical analysis of the activities, records and concept maps will be made. The results of this pilot application will be critically studied, aiming at an eventual reformulation before new applications.

Within the research project that supports this proposal, it is also being built a database of correlations[13] found with Google Correlate, which can serve as a basis to other teachers and researchers who may be interested in applying this proposal in their own classrooms.

**Conclusion**

We believe this is the first proposal to use Big Data tools acting as mediators in Science Teaching, with the computers as evocative objects (Turkle, 2007), the students thinking and learning *with-the-computer* (Rosa, 2008). It has an epistemological bias because it is not merely training in computational infrastructure or predictive analytics, but aims at providing students, future Science professionals, a better understanding of the construction of physical knowledge, especially of the phenomenon, observation, measurement, physical laws, theory, and causality concepts, among others. Furthermore, it is expected they gain a preparation, both in technical and ethical terms, for the scientific challenges posed by Big Data to the real world in which they will exercise their professions.

**References**

Ackermann, E. K. (2001). Piaget's Constructivism, Papert's Constructionism: What's the difference? Future of learning group publication, 5(3):438.

Baram-Tsabari, A.; Segev, E. (2009a). Just Google it! Exploring New Web-based Tools for Identifying Public Interest in Science and Pseudoscience. Proceeding of Chais Conference on Instructional Technologies Research 2009: Learning In The Technological Era, Feb. 18, Raanana, Israel, pp: 20-28.

---
[13] http://www.searchcorrelations.com

Baram-Tsabari, A.; Segev, E. (2009b). Exploring new web-based tools to identify public interest in science. Public Understanding of Science, 20(1):130-143.

Baram-Tsabari, A.; Segev, E. (in press). The half-life of a "teachable moment": The case of Nobel laureates. Public Understanding of Science.

Bülbül, M. Ş. (2009). Google Centered Search Method in Pursuit of Trends and Definitions in Physics and Education. [Online] Available: http://www.fizikli.com/piwi/fizikli6.pdf (February 7, 2014).

Butler, D. (2013). When Google got flu wrong. Nature, 494(7436):155–156.

Chalmers, David J. (2006). Strong and weak emergence. In: Clayton, Philip; Davies, Paul. The reemergence of emergence. New York: OUP - Oxford University Press. pp 244-256.

Dumbill, E.; Liddy, E. D.; Stanton, J.; Mueller, K.; Farnham, S. (2013). Educating the Next Generation of Data Scientists. Big Data, 1(1):21–27.

Field, H. (2003). Causation in a Physical World. In: M. J. Loux & D. W. Zimmerman (Eds.) Oxford Handbook of Metaphysics. Oxford: Oxford University Press.

Gantz, J.; Reinsel, D. (2012). The Digital Universe in 2020: Big Data, Bigger Digital Shadows, and Biggest Growth in the Far East. IDC - International Data Corporation, Framingham, MA. IDC 1414_v3.

Google Inc. (2013). Google Inc. Announces First Quarter 2013 Results. Mountain View, CA. [Online] Available: http://investor.google.com/earnings/2013/Q1_google_earnings.html (May 8, 2013).

IBM. (n.d.). What is big data? [Online] Available: http://www-01.ibm.com/software/data/bigdata/ (May 10, 2013).

Lazer, D.; Kennedy, R.; King, G.; Vespignani, A. (2014). The Parable of Google Flu: Traps in Big Data Analysis. Science, 343(6167):1203-1205.

Mattmann, C. A. (2013). Computing: A vision for data science. Nature, 493(7433):473-475.

Mayer, M. Yes, we are still all about search.[Blog post] Available: http://googleblog.blogspot.com.br/2006/05/yes-we-are-still-all-about-search.html (April 18, 2014).

Mohebbi, M. H.; Vanderkam, D.; Kodysh, J.; et al. (2011). Google Correlate Whitepaper. [Online] Available: http://www.google.com/trends/correlate/whitepaper.pdf (April 28, 2013).

Mohebbi, M. H. (2012). Google Correlate expands to 49 additional countries. [Blog post] Available: http://googleresearch.blogspot.com.br/2012/01/google-correlate-expands-to-49.html (April 6, 2014)

Novak, J. D.; Gowin, D. B. (1984). Learning how to Learn. Cambridge, MA: Cambridge University Press.

Papert, S. (1980). Mindstorms - Children, Computers and Powerful Ideas. New York: Basic Books.

Rosa, M. (2008). Constructing Identities through online Role Playing Game: relationships with the teaching and learning of mathematics in a distance learning course. Ph.D. Thesis, UNESP - São Paulo State University. Rio Claro, SP, Brazil.

Searls, D. (2013). People will do more with Big Data than big companies can [Blog post]. Available: http://blogs.law.harvard.edu/doc/2013/05/01/people-will-do-more-with-big-data-than-big-companies-can/ (May 7, 2013).

Segev, E.; Baram-Tsabari, A. (2012). Seeking science information online: Data mining Google to better understand the roles of the media and the education system. Public Understanding of Science, 21(7):813-829.

Sullivan, D. (2012). Google: 100 Billion Searches Per Month, Search To Integrate Gmail, Launching Enhanced Search App For iOS. [Online] Available: http://searchengineland.com/google-search-press-129925 (May 8, 2013).

Tufte, E. R. (2006). The Cognitive Style of PowerPoint: Pitching Out Corrupts Within. Cheshire, CT: Graphics Press.

Turkle, S. (2007). Evocative objects: things we think with. Cambridge, MA: MIT Press.

Ulbra. (2014). History and Epistemology of Physics Course Plan. Undergraduate Physics Program. Canoas, RS.

Vitinskii, Y. I. (1965). Solar Activity Forecasting. NASA, Washington, DC, TTF-289.

Yin, C.; Sung, H.-Y.; Hwang, G.-J.; et al. (2013). Learning by Searching: A Learning Environment that Provides Searching and Analysis Facilities for Supporting Trend Analysis Activities. Journal of Educational Technology & Society, 16(3):286-300.

Zikopoulos, P. C.; Deroos, D.; Parasuraman, K.; et al. (2013). Harness the Power of Big Data: The IBM Big Data Platform. New York: McGraw-Hill.